\documentclass[twocolumn,superscriptaddress]{revtex4-2}

\usepackage{amsmath, amssymb, amsfonts, geometry, physics, graphicx, hyperref}
\usepackage{geometry}
\usepackage{bm}
\usepackage{tensor}
\usepackage{braket}
\usepackage{gensymb}
\usepackage{color}
\usepackage{bbold}
\usepackage{enumitem}
\usepackage{tcolorbox}
\usepackage[mathscr]{euscript}
\usepackage[scr = mtc]{mathalfa}
\usepackage{booktabs}

\definecolor{Green}{RGB}{0,170,0}
\definecolor{Purple}{RGB}{102,0,255}
\definecolor{Blue}{RGB}{51,0,200}
\definecolor{Red}{RGB}{175,020,020}

\DeclareMathAlphabet\mathbfcal{OMS}{cmsy}{b}{n}

\newcommand{\spinthree}{\stackrel{\tiny \rm (3)}{\omega}}
\newcommand{\spinfour}{\stackrel{\scriptscriptstyle \rm (4)}{\omega}}

\newcommand{\polfourfree}{\stackrel{\scriptscriptstyle (4)}{\mathbb p}_{\text{\tiny{free}}}}
\newcommand{\polfourfreevec}{\stackrel{\scriptscriptstyle (4)}{\pmb{\mathbb p}}_{\text{\tiny{free}}}}

\newcommand{\polfourfixedvec}{\stackrel{\scriptscriptstyle (4)}{\pmb{\mathbb p}}_{\text{\tiny{fixed}}}}

\newcommand{\polthreefreevec}{\stackrel{\scriptscriptstyle (3)}{\pmb{\mathbb p}}_{\text{\tiny{free}}}}

\newcommand{\polAthreefixed}{\stackrel{\scriptscriptstyle (3)}{\pmb{\mathbb A}}_{\text{\tiny{fixed}}}}
\newcommand{\polBthreefixed}{\stackrel{\scriptscriptstyle (3)}{\pmb{\mathbb B}}_{\text{\tiny{fixed}}}}

\newcommand{\polthreefixedvec}{\stackrel{\scriptscriptstyle (3)}{\pmb{\mathbb p}}_{\text{\tiny{fixed}}}}

\newcommand{\polthreecircvec}{\stackrel{{\text{\tiny{(px)}}}}{\pmb{\mathbb m}}}

\newcommand{\wE}{{\omega_{\!\scriptscriptstyle E}}}
\newcommand{\xE}{\stackrel{\scriptscriptstyle \rm E}{x}}
\newcommand{\xEvec}{\stackrel{\scriptscriptstyle \rm E}{{\bf x}}}

\newcommand{\xL}{\stackrel{\scriptscriptstyle \rm L}{x}}

\newcommand{\khatvec}{{\hat{\pmb k}}}

\newcommand{\ksurf}{\stackrel{\scriptscriptstyle \approx}{k}}
\newcommand{\ksurfvec}{\stackrel{\scriptscriptstyle \approx}{\pmb k}}

\newcommand{\kfour}{k}
\newcommand{\kfourvec}{\pmb k}

\newcommand{\Afourvec}{\stackrel{\scriptscriptstyle (4)}{{\bf A}}}
\newcommand{\afour}{\stackrel{\scriptscriptstyle (4)}{a}}
\newcommand{\afourvec}{\stackrel{\scriptscriptstyle (4)}{{\bf a}}}

\newcommand{\chinumerical}{\stackrel{{\text{\tiny{(num)}}}}\chi}
\newcommand{\chianal}{\stackrel{{\text{\tiny{(anal)}}}}\chi}

\newcommand{\Tetrad}{e}

\newcommand{\Uvec}{{\pmb U}}
\newcommand{\uvec}{{\pmb u}}
\newcommand{\vvec}{{\pmb v}}

\newcommand{\mFWE}{\stackrel{{\text{\tiny{(FWE)}}}}{\pmb{\mathbb m}}}

\newcommand{\betavec}{\boldsymbol{\beta}}

\newcommand{\Liebeta}{\mathcal{L}_{\boldsymbol {\scriptscriptstyle\beta}} }
\newcommand{\LieU}{\mathcal{L}_{\bf {\scriptscriptstyle U}} }

\newcommand{\Gammathree}{\stackrel{\scriptscriptstyle \rm (3)}{\Gamma}}

\newcommand{\aAffinevec}{\pmb a}

\newcommand{\Dthree}{\mathcal{D}}
\newcommand{\Dthreebold}{\mathbfcal{D}}
\newcommand{\Dfour}{\stackrel{\scriptscriptstyle (4)}{\mathcal{D}}}
\newcommand{\gradfour}{\boldsymbol{\nabla}}

\newcommand{\eonevec}{{\pmb e}_{\hat 1}}
\newcommand{\etwovec}{{\pmb e}_{\hat 2}}
\newcommand{\ethreevec}{{\pmb e}_{\hat 3}}
\newcommand{\ezerovec}{{\pmb e}_{\hat 0}}
\newcommand{\eavec}{{\pmb e}_{\hat a}}
\newcommand{\ebvec}{{\pmb e}_{\hat b}}

\newcommand{\eivec}{{\pmb e}_{\hat i}}
\newcommand{\ejvec}{{\pmb e}_{\hat j}}

\newcommand{\eqvec}{{\pmb e}_{\hat q}}
\newcommand{\emuvec}{{\pmb e}_{\hat \mu}}

\newcommand{\eone}{e_{\hat 1}}
\newcommand{\etwo}{e_{\hat 2}}

\newcommand{\ea}{e_{\hat a}}
\newcommand{\eb}{e_{\hat b}}

\newcommand{\emu}{e_{\hat \mu}}

\newcommand{\ebetaonevec}{{\pmb e}_{\grave 1}}
\newcommand{\ebetatwovec}{{\pmb e}_{\grave 2}}

\newcommand{\ebetaone}{e_{\grave 1}}
\newcommand{\ebetatwo}{e_{\grave 2}}

\newcommand{\einertonevec}{{\pmb e}_{\breve 1}}
\newcommand{\einerttwovec}{{\pmb e}_{\breve 2}}

\newcommand{\einertavec}{{\pmb e}_{\breve a}}

\newcommand{\einertbvec}{{\pmb e}_{\breve b}}

\newcommand{\einertmuvec}{{\pmb e}_{\breve \mu}}

\hypersetup{
    colorlinks=true,
    linkcolor=blue,
    filecolor=magenta,
    urlcolor=cyan,
    citecolor=magenta
}

\begin{document}

\title{An Analytical Formula for Gravitational Faraday Rotation\\in the ADM Split of Spacetime}
\author{Mark T. Lusk}
\affiliation{Department of Physics, Colorado School of Mines, Golden, CO 80401, USA}
\date{\today}

\begin{abstract}
An analytical expression is derived for the rate of gravitational Faraday rotation measured by Eulerian observers. The reference frame is a Fermi-Walker triad aligned with the spatial wave vector.  Attention is restricted to the ADM split of Kerr spacetime and geometric optics. Our exact, closed-form GFR formula is implemented and verified to be consistent with numerical predictions.  The approach offers a new perspective on Faraday rotation, and it allows a single Eulerian observer to compare experimentally measured polarization holonomy with analytical prediction. Sliced spacetime does not suffer from a mathematical singularity at the ergosphere associated with Boyer-Lindquist coordinates in the threading decomposition. These physically intuitive coordinates can therefore be used to analytically produce and study GFR predictions for transits of light that pierce the ergosphere. 

\end{abstract}

\maketitle

\section{Introduction}\label{section:Intro}

Gravitomagnetic frame-dragging\cite{LenseThirring1918,LenseThirring1984} underlies the perspective that photon polarization is rotated by a spinning black hole\cite{Plebanski_1960, MTW}.  The effect has been theoretically studied for decades\cite{Plebanski_1960, Connors_1977, Chandrasekhar1998, Ishihara1988, FrolovShoom2011} by referencing the rotation to local Fermi-Walker frames aligned with the photon trajectory. This can be viewed as the rotation perceived by a Lagrangian ghost that accompanies the photon along its geodesic. 

In the present work, an alternative perspective is taken in which Fermi-Walker frames along the photon trajectory are each aligned with the \emph{local spatial wave vector}. The approach utilizes the  Arnowitt, Deser and Misner (ADM) \cite{ADM, Gourgoulhon2012} decomposition of spacetime into a series of three-dimensional space-like hypersurfaces. Eulerian observers\cite{Thorne1982} move normal to these foliations while preserving zero angular momentum.  We can then ask what rotation rate, with respect to the  alternative Fermi-Walker frames, would be perceived by Eulerian spotters stationed along the photon trajectory. 
 
Attention is restricted to Kerr spacetime\cite{Kerr2009} using Boyer-Lindquist coordinates \cite{BoyerLindquist1967}. Maxwell's equations are subjected to a geometric optics approximation, and a gauge is chosen such that polarization lies within the hypersurface, orthogonal to the spatial wave vector. For specified Killing parameters, analytical solutions can be obtained for the photon trajectory\cite{Gralla_2020, Lusk_PRD_2024, Lusk_Parvin_2025}. 

A locally orthonormal spacetime frame, dubbed the \emph{Shift Tetrad}, is constructed from the hypersurface normal, the tangent to the spatial wave vector, and a carefully constructed dyad of screen vectors. A closed-form, exact expression is then derived for the rate of gravitational Faraday rotation (GFR) in terms of a single component of the extrinsic curvature in this frame.  

The approach offers a way of clearly seeing how GFR is the result of hypersurface distortion, and the simple rate expression should be especially useful for closed trajectories, where there is a GFR holonomy that can be experimentally measured\cite{Lusk_Parvin_2025}.  

In addition, the Eulerian formulation of GFR does not suffer from a mathematical singularity at the ergosphere that is present for the Boyer-Lindquist coordinates in the threading decomposition of spacetime\cite{FrolovShoom2011}. This makes it possible to analytically produce and study GFR predictions for transits of light that pierce the ergosphere. 

The rate relation is implemented and shown to be consistent with numerical predictions for representative photon trajectories: a closed trajectory that stays outside the ergosphere; and a scattering path that goes in and out of the ergosphere as it arcs around the singularity.

\section{The ADM Formalism}\label{section:ADM}

\subsection{Introduction}

In ADM (\emph{3+1} or \emph{slicing}) decomposition, the evolution of hypersurfaces is characterized by \emph{scalar lapse function}, $\alpha$, and \emph{shift vector}, $\betavec$. The former quantifies the proper time interval between adjacent hypersurfaces. The shift vector describes the relative displacement of spatial points between slices, showing how the spatial coordinates on a foliation shift along coordinate time. Eulerian observers move orthogonal to the slices with 4-velocity $U^\mu$, the unit normal to the evolving hypersurface, and their congruence has zero vorticity.

A general spacetime metric can be expressed in matrix form as
\begin{equation}
[g_{\mu\nu}] = \left[
\begin{array}{cc}
 \left(-\alpha^2+\beta_k\beta^k\right) & \beta_j \\
 \beta_i  & \gamma_{ij} \\
\end{array}
\right]
\; .
\end{equation}
Greek letters indicate spacetime components with Latin used for their spatial counterparts. ${\boldsymbol\gamma}$ is the spatial metric. The metric basis vector $(\partial_t)^\mu$ is related to the normal and the shift vectors via
\begin{equation}\label{Ucontra}
U^\mu = \frac{1}{\alpha} \{ 1, -\beta^i \} 
\; .
\end{equation}
Since $\Uvec\cdot\Uvec = -1$, this implies that
\begin{equation}\label{Ucov}
U_\mu = \{-\alpha ,0,0,0 \}
\; .
\end{equation}

\subsection{Hypersurface Projection Operator}\label{subsection:Projs}

Define an operator, ${\bf h}$, that orthogonally projects vectors into the hypersurface:
\begin{equation}\label{hproj:defn:1}
h^{\mu}{}_{\nu} := g^{\mu}{}_{\nu} + U^{\mu}U_{\nu} 
\; .
\end{equation}
It is the induced metric and the pullback of the intrinsic hypersurface metric as well as its first fundamental form. The extrinsic curvature, ${\bf K}$, is the second fundamental form, and both figure prominently in what follows.   For any vector $V^\mu$, the projection $h^\mu{}_{\nu} V^\nu$ lies entirely within the spatial slice.

\subsection{Derivatives}\label{subsection:derivs}

For a spacetime manifold, the Lie derivative can be defined independent of any metric as\cite{Gourgoulhon2012}
\begin{equation}\label{Lie0}
\mathcal{L}_{\scriptscriptstyle \uvec}\vvec:=  \uvec \vvec -  \vvec \uvec
\; .
\end{equation}
This allows us to consider rates of change across the manifold in terms of tetrad frames. 

Let $\gradfour$ be the covariant derivative compatible with the spacetime metric, $\bf g$. Likewise  identify $\Dthreebold$ as the covariant derivative compatible with the slice metric, ${\boldsymbol\gamma}$. Using the spacetime metric, Lie derivatives can then be expressed as
\begin{equation}\label{Lie}
\mathcal{L}_{\scriptscriptstyle \uvec}\vvec \,\equiv\,  g(\uvec \, , \, \gradfour) \vvec -  g(\vvec \, , \, \gradfour) \uvec
\; .
\end{equation}
We use $g(\cdot,\cdot)$ and $\gamma(\cdot,\cdot)$ to distinguish between spacetime and foliation inner products. The spacetime and spatial gradient operators are related by
\begin{equation}\label{Goug1}
\Dthree_\rho V^\alpha = h^\alpha{}_\mu  h^\sigma{}_\rho \nabla_\sigma V^\mu .
\end{equation}

\section{Tetrads}\label{section:Tetrads}

We will find it expedient to represent fields in terms of  locally orthonormal sets of tetrads whose inner products comprise a Minkowski matrix \cite{MTW, Carroll2004, Tanaka_1996}.  Attention will be limited to a family for which $\ezerovec \equiv \Uvec$, the hypersurface normal, and $\ethreevec \equiv \khatvec$, the direction of the photon wave vector projection within the hypersurface.  The remaining two vectors span the screen plane of the spatial wave vector. The use of hatted indices identifies the vectors (subscripted) or 1-forms (superscripted) of a \emph{Generic Tetrad} within this family. The basis vectors and 1-forms can be defined as linear combinations of the metric basis sets, $\partial_\nu$ and $dx^{\nu}$, so that
\begin{equation}\label{tetrads:0p1}
\emuvec = \emu{}^{\nu} \, \partial_\nu \quad , \quad {\pmb e}^{\hat \mu} = e^{\hat \mu}{}_{\nu} \, dx^{\nu} 
\; 
\end{equation}
and 
\begin{equation}\label{tetrad:defn}
 \left[ \Tetrad^{\hat \nu}{}_{\!\mu} \right]
 =
\left[
\begin{array}{cccc}
 -\alpha  & 0 & 0 & 0 \\
 0 & e^{\hat{1}}{}_1 & e^{\hat{1}}{}_2 & e^{\hat{1}}{}_3
   \\
 0 & e^{\hat{2}}{}_1 & e^{\hat{2}}{}_2 & e^{\hat{2}}{}_3
   \\
 0 & \hat{k}_1 & \hat{k}_2 & \hat{k}_3 \\
\end{array}
\right]
\; .
\end{equation}
The tetrad components of a vector are raised and lowered using the Minkowski metric, $\eta_{\hat{\mu}\hat{\nu}}$, while the spacetime metric tensor, $g_{\rho \sigma}$, facilitates raising and lowering of the metric indices (no hats). 

Vectors can be expressed in terms of their components in a generic orthonormal tetrad frame:
\begin{equation}\label{tetrad:Vectors}
V^{\hat \nu} = \Tetrad^{\hat \nu}{}_{\!\mu} V^\mu \; .
\end{equation}
The hypersurface is spanned by orthonormal triad, $\{ \eonevec, \etwovec, \khatvec \}$.

\section{Spacetime Spin Coefficients}\label{section:Spin:Spacetime}

Tetrad spin coefficients are defined by
\begin{equation}\label{spin:defn1}
\spinfour_{\hat{i}\hat{j}\hat{q}} 
:=    g\bigl(\eivec,\, \gradfour_{\eqvec}  \ejvec\bigr)
\equiv  
      e_{\hat i}{}^{p} e_{\hat q}{}^{\rho}\, \nabla_{\rho} e_{\hat j p} 
\, ,
\end{equation}
where we have used $\nabla_{\rho}g_{\ell p} = 0$.

The quantity $\gradfour_{\eqvec}  \ejvec \equiv (\eqvec , \gradfour) \ejvec$ measures the infinitesimal change of $\ejvec$ when transported along the direction $\eqvec$, a directional derivative. The component $\spinfour_{\hat{i}\hat{j}\hat{q}}$ projects this change onto $\eivec$, quantifying a differential rotation about $\ejvec$ of $\eqvec$ towards $\eivec$. For example, $\spinfour_{\hat{1}\hat{2}\hat{3}}$ corresponds to a right-handed rotation of the $\hat{1}$--$\hat{2}$ screen dyad about its spatial wave vector.

Antisymmetry with respect to the first two indices is immediate by application of the Leibniz rule to
\begin{equation}\label{RCC:antisymm}
0 = g(\eqvec , \gradfour) g(\eivec,\, \ejvec) 
\; .
\end{equation}
Expanding the derivative, and expressing the result in terms of spin coefficients, we have
\begin{equation}\label{spin:antisymm}
  \spinfour_{\hat{i}\hat{j}\hat{q}} = -\!\spinfour_{\hat{j}\hat{i}\hat{q}}  \; .
\end{equation}

\section{Extrinsic Curvature}\label{section:curvature}

 The extrinsic curvature of the hypersurface is defined as the Lie derivative of the projection tensor along the hypersurface normal\cite{Carroll2004}Appendix D, \cite{MTW}Ex.2.19, \cite{Poisson2004}\S 3.4:
\begin{equation}\label{curvatureLie}
{\bf K} = -\frac{1}{2} \mathcal{L}_{\scriptscriptstyle \Uvec}{\bf h} 
\; .
\end{equation}
Using the definition of the hypersurface normal, Eq. \ref{Ucontra}, this can be expanded out and then significantly reduced, because several inner products are between hypersurface vectors and its normal. Within our stationary spacetime setting, the extrinsic curvature has metric components of\cite{Carroll2004}
\begin{equation}\label{K:ADM}
K_{ij} \;=\; \frac{1}{2\alpha}\big(\Dthree_i\beta_j+\Dthree_j\beta_i\big)
\; .
\end{equation}

Extrinsic curvature can be equivalently defined using the Weingarten map between hypersurface tangent vectors and the gradient of the surface normal \cite{MTW}Eq. (21.60), \cite{Gourgoulhon2012}Eq. (3.20):
\begin{equation}\label{Kop2}
{\bf K}(\uvec\, , \, \vvec)  \equiv g(\uvec\, , \, {\bf K}\vvec) := -g(\uvec \, , \, \gradfour_\vvec \Uvec ) 
\; .
\end{equation}
Here $\uvec$ and $\vvec$ are arbitrary spatial vectors. Using the definition of spin coefficients, Eq. \ref{spin:defn1}, and exploiting their antisymmetry, Eq. \ref{spin:antisymm}, we then have
\begin{equation}\label{K:defn:3}
K_{\hat i \hat j} = \spinfour_{\hat 0\hat i\hat j}
\; .
\end{equation}
Multiply both sides by $e^{\hat i}{}_{m} e^{\hat j}{}_{n}$ and reduce to find that 
\begin{equation}\label{K:Tetrads}
K_{mn} = K_{\hat i \hat j}  e^{\hat i}{}_{\! m} e^{\hat j}{}_{\! n} 
\; .
\end{equation}
This can also be inverted:
\begin{equation}\label{K:Tetrads:v2}
K_{\hat m \hat n}  =  e_{\hat m}{}^{\!p} e_{\hat n}{}^{\!q}K_{pq} 
\;.
\end{equation}
%

\section{Spatial Spin Coefficients}\label{section:Spin:Hypersurface}

We can define hypersurface spin coefficients using only the spatial triad $\{\eivec\}$ and the intrinsic covariant derivative $\Dthreebold$ on $\Sigma_t$:
\begin{equation}\label{spin3D:1}
\spinthree_{\hat{i}\hat{j}\hat{q}} := \gamma\bigl(\eivec,\, \Dthreebold_{\eqvec}  \ejvec\bigr)  \; .
\end{equation}
These can be related to their spacetime counterparts using the Gauss-Weingarten equation \cite{Gauss1828, Weingarten1861, Poisson2004}:
\begin{equation}\label{Gauss:Weingarten}
\gradfour_{\eivec} \ejvec = \Dthreebold_{\eivec} \ejvec + K_{{\hat i}{\hat j}} \ezerovec
\; .
\end{equation}
Here $K_{{\hat i}{\hat j}} \,\ezerovec$ is the associated Weingarten map (shape operator) that represents the normal component of the derivative. Project this relation onto hypersurface vector, $\eqvec$. Since $g(\eqvec , \ezerovec) \equiv  g(\eqvec , \Uvec) = 0$, we have
\begin{equation}
g( \eqvec \, , \, \gradfour_{\eivec} \ejvec ) = \gamma( \eqvec \, , \, \Dthreebold_{\eivec} \ejvec ) 
\; .
\end{equation}
Using Eqs. \ref{spin:defn1} and \ref{spin3D:1}, and changing indices for ease of reading, this is equivalent to
\begin{equation}\label{RCC4andRCC3}
\spinfour_{{\hat i}{\hat j}{\hat q}} \,=\, \spinthree_{{\hat i}{\hat j}{\hat q}}
\; .
\end{equation}

\section{Parallel Transport of Polarization}\label{section:SpacetimeParallelTransport}

\subsection{Parallel Transport Equation}\label{section:Maxwell}

In the (3+1) slicing decomposition of a stationary spacetime, the photon frequency (energy) measured by Eulerian observers is identified as $\wE$. A harmonic time dependence of fields is therefore represented as $e^{i \wE t}$, and the Eulerian frequency is then identified as the rate of change of Eikonal phase, $S$, as perceived by an Eulerian observer:
\begin{equation}\label{wEdefn1}
\wE := -U_{\mu}\kfour^{\mu} \equiv  \alpha  \kfour^0 
\, .
\end{equation}

The spacetime Maxwell equations can then be subjected to an Eikonal ansatz and a perturbation expansion, using $\wE$ to identify terms of differing order. Within this geometric optics approximation, the wave vector and the polarization are both governed by the  parallel transport equation \cite{MTW, Lusk_PRD_2024}, which we briefly review. 

Assume a vacuum setting and consider vector potential $\Afourvec$ in the Lorenz gauge. The requisite short-wavelength ansatz is
\begin{equation}
\Afourvec(x)=\Re\{\afourvec(x)\,e^{i S(x) \wE}\},\quad  \wE \gg 1
\; ,
\end{equation}
with real phase, $S$, and complex amplitude, $\afourvec$, assumed to be slowly varying. 

The \emph{spacetime wave 1-form} is defined in terms of derivatives of the Eikonal phase, S:
\begin{equation}\label{kfourAsPhaseGradient}
\kfour_\mu =\nabla_{\mu}S\equiv \partial_{\mu}S
\; .
\end{equation}
Insert this into Maxwell’s equations and collect powers of $\wE$.  Define the \emph{affine parameter, $\lambda$}, as the unique parameterization of the trajectory such that
\begin{equation}\label{affinedefn}
\kfour^\mu  = \frac{dx^\mu }{d\lambda} .
\end{equation}
At leading order ($\wE^{2}$) in the Eikonal expansion, the wave vector is null valued:
\begin{equation}
g(\kfourvec , \kfourvec ) =0 
\; .
\end{equation}
Using $k^\mu = g^{\mu\nu} k_\nu$, we obtain two useful relations between the contravariant and covariant components of the spacetime wave vector:
\begin{align}\label{kcontracov}
\kfour^i =& \gamma^{ij}\kfour_j-\frac{\wE}{\alpha}\beta^i \nonumber \\
\kfour_i  =& \gamma_{ij}\kfour^j + \frac{\wE}{\alpha}\beta_i
\, .
\end{align}

After applying the Lorenz gauge, $(\kfourvec ,\afourvec)=0$, the next order in the Eikonal expansion ($\wE^{1}$)  produces a transport equation for the field amplitude:
\begin{equation}
(\kfourvec,\gradfour) \afourvec + \tfrac12 \afourvec (\gradfour\cdot \kfourvec) = 0
\; .
\end{equation}
Expand and simply this equation after substituting in the unit-valued polarization,
\begin{equation}\label{pol:defn:0}
\afourvec \,=: \,\afour\polfourfreevec 
\; ,
\end{equation}
to obtain the parallel transport equation:
\begin{equation}\label{pol:evol}
\boxed{
g(\kfourvec,\gradfour)  \polfourfreevec = 0
}
\; . 
\end{equation}
This can be expressed in terms of metric components using Eq. \ref{affinedefn}:
\begin{equation}\label{PX4:1}
\frac{d\polfourfree^\mu}{d\lambda} + \Gamma^\mu{}_{\!\nu \rho} \polfourfree^\rho \kfour^\nu = 0 
\; .
\end{equation}
The subscript \emph{free} serves as a reminder that the polarization gauge freedom has not yet been fixed.

\section{Spatial Wave Vector $\ksurfvec$}

Define the \emph{spatial wave vector}, $\ksurfvec$, as
\begin{equation}
\ksurfvec := {\bf h}\kfourvec \equiv h \gradfour S \neq \Dthreebold S,
\end{equation}
where the projection operator, ${\bf h}$, is given in Eq. \ref{hproj:defn:1}. 
It has foliation adapted coordinates of
\begin{equation}\label{ksurfcontra}
\ksurf^i := k^i + \frac{\wE }{\alpha}\beta^i 
\; .
\end{equation}
Recalling Eq. \ref{affinedefn}, the spatial wave vector can be also be expressed as
\begin{equation}\label{ksurfcontra1}
\ksurf^i = \frac{dx^i }{d\lambda} + \frac{\wE }{\alpha}\beta^i \,\equiv\, \frac{dx^i }{d\lambda} - \wE U^i 
\; ,
\end{equation}
where the Eulerian frequency is given in Eq. \ref{wEdefn1}. 

Frequency $\wE$ accounts for time dilation (red/blue shift), so we see that the spatial wave vector, $\ksurfvec$, is the difference between the spatial components of the spacetime photon trajectory and the dilation-corrected velocity of the Eulerian observer. In other words, $\ksurfvec$, is tangent to the spatial photon trajectory as perceived by the Eulerian observer. It can be viewed as the derivative of a parametrized position, $\xEvec\!(\lambda)$, along the photon trajectory as viewed by the Eulerian observer:
\begin{equation}\label{xExL:1}
\frac{d\!\xE^i}{d\lambda} := \frac{d\!\xL^i }{d\lambda} - \wE U^i \,\equiv\,\,  \ksurf^i .
\end{equation}
Here $\xL\!\!(\lambda)$ is the Lagrangian trajectory obtained by simply selecting the spatial components of the parametrized spacetime trajectory, $x^\mu(\lambda)$. 

The ability to represent the spatial wave vector as a rate of change of spatial position, Eq. \ref{xExL:1}, is of central importance to us, because we will encounter spatial covariant directional derivatives of the form
\begin{align}\label{surftang6}
\gamma( \ksurfvec \, , \, \Dthreebold )V^i 
=&\frac{d\!\xE^j}{d\lambda}  \left( \frac{\partial V^i}{\partial x^j} + \Gammathree^i{}_{\!\!\!j \ell} V^\ell \right) \nonumber \\
=& \frac{d V^i}{d\lambda}   + \Gammathree^i{}_{\!\!\!j \ell} V^\ell \ksurf^i
\; .
\end{align}
The combination of Eqs. \ref{xExL:1} and \ref{surftang6} gives the following important representation of such directional derivatives:
\begin{equation}\label{DirecDerivEulerian}
\gamma(\ksurfvec \, , \, \Dthreebold )V^i =  \frac{d V^i}{d\lambda}   +  \Gammathree^i{}_{\!\!\!j \ell} V^\ell  \ksurf^j 
\; .
\end{equation}

Using the fact that the spacetime wave vector is null-valued, it is straightforward to show that the magnitude of the spatial wave vector is $\wE$.  A normalized projected wave vector, $\khatvec$, is therefore
\begin{equation}\label{khat}
\khatvec \,:=\,  \tfrac{1}{\wE}\ksurfvec 
\; .
\end{equation}
This is the direction the photon is seen to move, within the hypersurface, from the perspective of an Eulerian observer. The latter moves orthogonal to the hypersurface at a velocity of $U^\mu$, while the photon wave vector is $\kfour^\mu$, which is tangent to the path of light. The trajectory of the photon relative to the observer is therefore the difference between these two.

\section{Gauge-Modified Parallel Transport of Polarization}\label{section:GaugeModPX}

Spacetime photon polarization can be projected onto the hypersurface, using Eq. \ref{hproj:defn:1}, to construct an intrinsic hypersurface polarization vector:
\begin{equation}\label{pol3free}
\polthreefreevec := {\bf h} \polfourfreevec 
\; .
\end{equation}
This is the polarization measured by an Eulerian observer, identified as \emph{free} because the polarization gauge has not yet been fixed. 

The gauge freedom is due to the null character of the wave vector, which allows a weighted wave vector to be added to the polarization without changing their inner product:
\begin{equation}\label{pol4fixed:defn}
\polfourfixedvec =\polfourfreevec + \,\xi \kfourvec
\; .
\end{equation}
Here $\xi$ is a scalar gauge function that is chosen to satisfy an imposed constraint. In the present work, the gauge is fixed by insisting  that the polarization be orthogonal to both the hypersurface normal and the spatial wave vector. (The temporal component is then equal to zero.)  This gauge-fixed polarization satisfies the Polarization-Fixed Evolution equation: 
\begin{equation}\label{gauge:GMPX}
\boxed{
g(\kfourvec,\gradfour )\polfourfixedvec = \kfourvec \frac{d\xi}{d\lambda} 
}
\; .
\end{equation}

Consistent with the free hypersurface polarization of Eq. \ref{pol3free},  define the gauge-fixed hypersurface polarization as
\begin{equation}\label{pol3fixed}
\polthreefixedvec := {\bf h} \polfourfixedvec \equiv \, \polthreefreevec + \, \xi \ksurfvec\
\; .
\end{equation}
The spatial wave vector, $\ksurfvec$, is defined in Eq. \ref{ksurfcontra}. Fix the gauge function, $\xi$, by insisting that the hypersurface polarization be orthogonal to the spatial wave vector (and also to the hypersurface normal by construction):
\begin{equation}
\gamma(\polthreefixedvec , \ksurfvec) = 0 \; . 
\end{equation}
Expanding this using Eq. \ref{pol3fixed} gives the requisite gauge function as 
\begin{equation}\label{gauge:function}
\xi  = \frac{\gamma\bigl({\bf h} \polfourfreevec  , \ksurfvec \bigr)}{\wE^2}
\; ,
\end{equation}
and the resulting spatial polarization is
\begin{equation}\label{gauge:polfixed}
\polthreefixedvec = {\bf h} \polfourfreevec  +\frac{\gamma\bigl({\bf h} \polfourfreevec  , \ksurfvec \bigr)}{\wE^2} \ksurfvec 
\; .
\end{equation}

\section{Fermi-Walker Transport}\label{section:FWE}

The evolution of spatial polarization can be referenced with a spatial Fermi-Walker frame\cite{MTW}\S 6.5 that is instantaneously aligned with the spatial wave vector, $\ksurfvec$, and not the trajectory of the photon. Every spatial point along the photon trajectory can host such a frame, and each exhibits a rotation rate about the spatial wave vector that is the minimum required to maintain orthogonality to it. An analogous frame for Lagrangian ghosts is aligned with the photon trajectory\cite{Lusk_PRD_2024, Lusk_Parvin_2025}, so the two types of observers use distinct Fermi-Walker reference frames to quantify polarization. The reference frame dynamics will be identified as FWE (Eulerian) and FWL (Lagrangian).

FWE dynamics is best expressed in terms of the \emph{absolute spatial derivative} of its basis vectors, $\eivec$, which amounts to a re-expression of the directional derivative of Eq. \ref{DirecDerivEulerian} in which $\ksurfvec$ is identified as a rate using Eqs. \ref{xExL:1} and \ref{khat}:
\begin{equation}\label{deriv:spatial:abs}
\frac{\Dthree \eivec}{d\lambda} = \wE \gamma(\khatvec  \, , \, \Dthree ) \eivec  
\; .
\end{equation}
A spatial vector $\bf V$ is said to be FWE along the spatial wave vector if 
\begin{equation}\label{FWX1}
\frac{\Dthree {\bf V}}{d\lambda}  = \gamma(\khatvec , {\bf V}) \aAffinevec -  \gamma(\aAffinevec , {\bf V}) \khatvec 
\; ,
\end{equation}
where $\aAffinevec$ is the affine acceleration of the unit spatial wave vector,
\begin{equation}\label{aAffine}
\aAffinevec := \frac{\Dthree \khatvec}{d \lambda} 
\; .
\end{equation}
We can now define the \emph{FW tetrad}, $\{ \einertmuvec \}$, a member of the Generic Tetrad family. Apply the FWE equation to a FW screen vector, $\einertbvec$, one of two unit vectors that span the plane orthogonal to $\khatvec$:
\begin{equation}\label{FWX2}
\frac{\Dthree \einertbvec}{d\lambda} = \gamma(\khatvec \, , \, \einertbvec) \aAffinevec - \, \gamma(\aAffinevec \, , \, \einertbvec) \khatvec  
\; .
\end{equation}
But $\gamma(\khatvec , \einertbvec) = 0$, and $\gamma(\aAffinevec \, , \, \einertbvec) = \wE \!\!\spinthree_{\breve{b}\breve{3}\breve{3}}$, so we have
\begin{equation}\label{FWE:spina33}
\boxed{
\frac{\Dthree \einertbvec}{d\lambda} =  -\wE\spinthree_{\breve{b}\breve{3}\breve{3}}\khatvec  \, , \quad b\in\{1, 2 \}
}
\; .
\end{equation}

\section{Rate of GFR}\label{section:GFR}

The evolution of polarization is governed by Eq. \ref{gauge:GMPX}, and the gauge fixing of Eq. \ref{gauge:polfixed} implies that spatial polarization can be expressed as a linear combination of the orthonormal screen vectors of our Generic Tetrad:
\begin{equation}\label{pol3:defn}
\polthreefixedvec = \eonevec \cos\zeta\ + \etwovec \sin\zeta
\; .
\end{equation}
Here $\zeta(\lambda)$ is defined as the polarization vector angle with respect to the Generic Screen Dyad $\{ \eonevec, \etwovec \}$, with $\zeta(0) = 0$.  

Analogous to the absolute spatial derivative of FWE, it is useful to express the evolution equation in terms of the \emph{absolute spacetime derivative} of a tetrad basis vector, $\emuvec$, defined as
\begin{equation}\label{deriv:spacetime:abs}
\frac{\Dfour \emuvec}{d\lambda} = g(\kfourvec \, , \, \gradfour) \emuvec     
\; .
\end{equation}
Here the wave vector, $\kfourvec$, is related to the affine parameterization by Eq. \ref{affinedefn}.

\subsection{GFR in the Generic Frame}\label{subsection:GFR:betaTetrad:generic}

Use Eq. \ref{deriv:spacetime:abs} to take the absolute derivative of the polarization of Eq. \ref{pol3:defn}:
\begin{equation}
\frac{\Dfour}{d\lambda} \polfourfixedvec
= \frac{\Dfour}{d\lambda}(\eonevec \cos\zeta)  + \frac{\Dfour}{d\lambda} ( \etwovec \sin\zeta ) 
\; .
\end{equation}
Expand the derivative terms, project the result onto $\eonevec$, and simplify to obtain
\begin{align}\label{pol:derivproj}
g\biggl(\eonevec ,& \frac{\Dfour}{d\lambda} \polthreefixedvec \biggr) \nonumber \\
&=
   - \frac{d\zeta}{d\lambda}\sin\zeta  
  \,+\, 
   g\biggl(\eonevec , \frac{\Dfour\etwovec}{d\lambda}\biggr) \sin\zeta 
   \; .
\end{align}
Project the polarization evolution of Eq. \ref{gauge:GMPX} onto $\eonevec$, and replace the left-hand side with Eq. \ref{pol:derivproj} to find that
\begin{equation}
   - \frac{d\zeta}{d\lambda}\sin\zeta  
+ 
   g\biggl(\eonevec , \frac{\Dfour\etwovec}{d\lambda}\biggr) \sin\zeta = 0 
   \; .
\end{equation}
Cancellation of the common $\sin\zeta$ terms gives
\begin{equation}\label{chirate:Tetrads}
\frac{d\zeta}{d\lambda}  
= g\biggl(\eonevec ,\frac{\Dfour\etwovec}{d\lambda}\biggr) 
\; .
\end{equation}
Express the wave vector in terms of the tetrad basis, then use Eqs. \ref{deriv:spacetime:abs} and \ref{spin:defn1}  to re-write Eq. \ref{chirate:Tetrads} in terms of spin coefficients. From Eq. \ref{khat}, we have 
\begin{equation}
\kfour^{\hat 0} = \kfour^{\hat 3} = \wE, \quad \kfour^{\hat 1} = \kfour^{\hat 2} = 0 
\; .
\end{equation}
These can be used with Eq. \ref{tetrad:Vectors} to express the spacetime wave vector in terms of the generic basis. Eq. \ref{chirate:Tetrads} then reduces to
\begin{equation}\label{rate:zeta:final}
\boxed{
\frac{d\zeta}{d \lambda}  = \wE \bigl(\spinfour_{\hat{1}\hat{2}\hat{0}} \,+\, \spinfour_{\hat{1}\hat{2}\hat{3}} \bigr)  
}
\; .
\end{equation}

\subsection{Tetrad Rotation in the FW Frame}\label{subsection:Tetrad:rotrate}

With a nascent expression for the rate of GFR in hand, we turn attention to the angle, $\varphi$, by which the FW frame dyad, $\{ \einertavec \}$, is rotated relative to the Generic Screen Dyad:
\begin{align}\label{FW:dyad}
\einertonevec =& \cos\!\varphi\,\eonevec - \sin\!\varphi\,\etwovec \nonumber \\
\einerttwovec =& \sin\!\varphi\,\eonevec + \cos\!\varphi\,\etwovec
\; .
\end{align}
The FW screen dyad is governed by FWE, Eq. \ref{FWE:spina33}, so
\begin{equation}\label{FWE:spina33:v2}
\frac{\Dfour\einertonevec }{d\lambda}= -\wE\spinthree_{\breve{1}\breve{3}\breve{3}}\khatvec 
\; .
\end{equation}
Substitute in our Eq. \ref{FW:dyad} expression for $\eonevec$ to obtain
\begin{equation}
\frac{\Dfour \left( \cos\!\varphi\,\eonevec - \sin\!\varphi\,\etwovec\right)\pmb{ } }{d\lambda}
= 
-\wE\spinthree_{\breve{1}\breve{3}\breve{3}}\khatvec 
\; .
\end{equation}
Expand the derivatives, re-arrange, and project the result onto $\eonevec$:
\begin{align}
\frac{d\varphi}{d\lambda} (-\sin\!\varphi) &
+ \cos\!\varphi\,\left(\eonevec,\frac{\Dfour \eonevec}{d\lambda}\right) \nonumber \\
&- \sin\!\varphi\,\left(\eonevec,\frac{\Dfour \etwovec}{d\lambda}\right)
=0
\; .
\end{align}
Since
\begin{equation}
\biggl(\eonevec,\frac{\Dfour\eonevec}{d\lambda}\biggr) = 0 
\; ,
\end{equation}
our equation reduces to
\begin{equation}
\frac{d\varphi}{d\lambda} (-\sin\!\varphi) - \sin\!\varphi \, \biggl(\eonevec,\frac{\Dfour \etwovec }{d\lambda} \biggr)
= 0
\; .
\end{equation}
Cancel out the common sine term to obtain the rate of rotation of the FW frame with respect to the Generic Tetrad. Eqs. \ref{RCC4andRCC3} and \ref{deriv:spatial:abs} allow the result to be expressed in terms of a spin coefficient:
\begin{equation}\label{rate:varphi:1}
\boxed{
\frac{d\varphi}{d\lambda}  = -\wE \spinfour_{\hat{1}\hat{2}\hat{3}}  
}
\;.
\end{equation}

\subsection{GFR in the FW Frame}\label{subsection:GFR:inert} 

Eq. \ref{rate:zeta:final} gives the rate of GFR with respect to our Generic Tetrad, and the rate of rotation of the Generic Tetrad with respect to the FW frame is given by the negative of Eq. \ref{rate:varphi:1}. Putting the two rates together, we obtain a GFR rate  relation for the angle, $\chi$, through which polarization rotates with respect to the FW frame:
\begin{equation}\label{rate:chi:1}
\boxed{
\frac{d\chi}{d\lambda} := \frac{d(\zeta + \varphi)}{d\lambda} = \wE \spinfour_{\hat{1}\hat{2}\hat{0}} 
}
\; .
\end{equation}
This result holds for any member of the Generic Tetrad family.

\subsection{Evaluation of $\spinfour_{{\hat 1}{\hat 2}{\hat 0}}$}\label{subsection:spinab0}

The Lie derivative defined in Eq. \ref{Lie} implies that
\begin{equation}
\gradfour_{\scriptscriptstyle \Uvec} \ebvec = \LieU \ebvec + \gradfour_{\ebvec} \Uvec
\; ,
\end{equation}
and taking the inner product with $\eavec$ gives 
\begin{equation}\label{spin:ab0:2}
\spinfour_{\hat a\hat b\hat 0} \, = \, g(\eavec,\mathcal \LieU \ebvec) + g(\eavec,\gradfour_{\ebvec}\Uvec)
\; .
\end{equation}
Here and throughout, subscripts $\{ a, b \} \in \{ 1, 2 \}$. Using Eq. \ref{Kop2}, this can be re-expressed in terms of the extrinsic curvature:
\begin{equation}\label{spin:ab0:3}
\spinfour_{\hat a\hat b\hat 0}\, = \, g(\eavec,\mathcal \LieU \ebvec)  - K_{\hat a\hat b}
\; .
\end{equation}

The Lie derivative of lapse function, $\alpha$, is simply a directional derivative:
\begin{equation}\label{Lie:scalar}
\LieU \alpha = ( \Uvec , \gradfour \alpha)
\; .
\end{equation}
In Kerr spacetime, though, $\alpha = \alpha(r, \theta)$, so its gradient only has components in the $r$ and $\theta$ directions. In contrast, the hypersurface normal has contravariant components that are zero in these two directions, evident in Eq. \ref{Ucontra} by noting that $\betavec$ is non-zero only in the $\phi$-direction. Therefore
\begin{equation}\label{Lie:alpha1}
\LieU \alpha = 0
\; .
\end{equation}
Using the definition of the hypersurface normal, Eq. \ref{Ucontra}, we then have
\begin{equation}
\LieU \ebvec = \frac{1}{\alpha}\big(\mathcal L_{\partial_t} \ebvec - \Liebeta \ebvec\big) \; .
\end{equation}
Substitute this into Eq. \ref{spin:ab0:3} and invoke the linearity of the inner product:
\begin{equation}\label{spin:ab0:4}
\spinfour_{\hat a\hat b\hat 0}
=\frac{1}{\alpha}\,g\big( \eavec ,\mathcal L_{\partial_t} \ebvec \big)
-\frac{1}{\alpha} \, g\big( \eavec, \Liebeta \ebvec \big)
- K_{\hat a\hat b}.
\end{equation}

None of the steps above would change if we had reversed the assignment of $\eonevec$ and $\etwovec$, so Eq. \ref{spin:ab0:4} can be written with the first two indices reversed:
\begin{equation}\label{spin:ab0:5}
\spinfour_{\hat b\hat a\hat 0}
=\frac{1}{\alpha}\,g\big( \ebvec ,\mathcal L_{\partial_t} \eavec \big)
-\frac{1}{\alpha} \, g\big( \ebvec, \Liebeta \eavec \big)
- K_{\hat b\hat a}.
\end{equation}
The spin antisymmetry of Eq. \ref{spin:antisymm} implies that
\begin{equation}
\spinfour_{\hat a\hat b\hat0}
=\frac{1}{2} \left( \spinfour_{\hat a\hat b\hat0} - \spinfour_{\hat b\hat a\hat0} \right)
\; .
\end{equation}
Inserting Eqs. \ref{spin:ab0:4} and \ref{spin:ab0:5} into the right side of this relation gives
\begin{align}
&\spinfour_{\hat a\hat b\hat0} \nonumber \\
&= \frac{1}{2\alpha} \left( g\big( \eavec ,\mathcal L_{\partial_t} \ebvec \big)
- g\big( \eavec, \Liebeta \ebvec \big)
- \alpha K_{\hat a\hat b}\right) \nonumber \\
& - \frac{1}{2\alpha} \left( g\big( \ebvec ,\mathcal L_{\partial_t} \eavec \big)
-g\big( \ebvec, \Liebeta \eavec \big)
- \alpha K_{\hat b\hat a} \right) 
\; .
\end{align}
Since the extrinsic curvature is symmetric, the curvature terms cancel leaving
\begin{align}\label{spin:ab0:6}
\spinfour_{\hat a\hat b\hat0}
&= \frac{1}{2\alpha} \left( 
g\big( \eavec ,\mathcal L_{\partial_t} \ebvec \big) 
\, - \,
g\big( \ebvec ,\mathcal L_{\partial_t} \eavec \big)
\right) \nonumber \\
& \,- \frac{1}{2\alpha} \left( 
\, g\big( \eavec, \Liebeta \ebvec \big) 
\, - \, 
g\big( \ebvec, \Liebeta \eavec \big) 
\right) 
\; .
\end{align}

We next show that the first pair of terms on right side of Eq. \ref{spin:ab0:6} is equal to zero. The orthonormality relation,
\begin{equation}
\gamma_{ij} \, e_{\hat a}{}^{i} \, e_{\hat b}{}^{j} = \delta_{\hat a \hat b}
\; ,
\end{equation}
can be differentiated with respect to $t$, giving
\begin{equation}
\partial_t\big(\gamma_{ij} e_{\hat a}{}^{i} e_{\hat b}{}^{j}\big) = 0
\; .
\end{equation}
Expand the derivative, with attention restricted to stationary spacetimes so that $ \partial_t\gamma_{ij} = 0$:
\begin{equation}\label{relation:1}
\gamma_{ij} \, ( \partial_t e_{\hat a}{}^{i})\,e_{\hat b}{}^{j}
+
  \gamma_{ij} \, e_{\hat a}{}^{i} ( \partial_t e_{\hat b}{}^{j} ) = 0
\; .
\end{equation}
It will be convenient to label these terms,
\begin{align}
T^{(2)}_{\hat a \hat b} &:= \gamma_{ij} \, (\partial_t e_{\hat a}{}^{i})\,e_{\hat b}{}^{j}, \\
T^{(3)}_{\hat a \hat b} &:= \gamma_{ij} \, e_{\hat a}{}^{i} \, (\partial_t e_{\hat b}{}^{j})
\; 
\end{align}
so that Eq. \ref{relation:1} is equivalent to
\begin{equation}\label{relation:2}
T^{(2)}_{\hat a \hat b} + T^{(3)}_{\hat a \hat b} = 0
\; .
\end{equation}
Since the indices can be reversed, we also have 
\begin{equation}\label{relation:3}
T^{(2)}_{\hat b \hat a} + T^{(3)}_{\hat b \hat a} = 0
\; .
\end{equation}
Subtract Eq. \ref{relation:3} from Eq. \ref{relation:2}:
\begin{equation}
0 = \tfrac12\big(T^{(2)}_{\hat a\hat b}-T^{(2)}_{\hat b\hat a}\big)
+\tfrac12\big(T^{(3)}_{\hat a\hat b}-T^{(3)}_{\hat b\hat a}\big)
\; .
\end{equation}
Noting that
\begin{align}\label{relation:4}
T^{(2)}_{\hat a\hat b}-T^{(2)}_{\hat b\hat a}
&= \gamma_{ij}\big(\partial_t e_{\hat a}{}^{i}\,e_{\hat b}{}^{j}
    -\partial_t e_{\hat b}{}^{i}\,e_{\hat a}{}^{j}\big) \nonumber \\
T^{(3)}_{\hat a\hat b}-T^{(3)}_{\hat b\hat a}
&= \gamma_{ij}\big(e_{\hat a}{}^{i}\,\partial_t e_{\hat b}{}^{j}
    - e_{\hat b}{}^{i}\,\partial_t e_{\hat a}{}^{j}\big)
\; ,
\end{align}
and adding these gives
\begin{equation}\label{relation:6}
\gamma_{ij}\,e_{[\hat a}{}^{i}\,\partial_t e_{\hat b]}{}^{j} = 0
\; .
\end{equation}
Because $(\mathcal L_{\partial_t}e_{\hat b})^{j}=\partial_t e_{\hat b}{}^{j}$ in the metric frame,
\begin{equation}\label{relation:7}
\gamma_{ij}\,e_{\hat a}{}^{i}\,(\mathcal L_{\partial_t} e_{\hat b})^{j}
= \gamma_{ij}\,e_{\hat a}{}^{i}\,\partial_t e_{\hat b}{}^{j}
\; .
\end{equation}
Apply Eq. \ref{relation:7} to Eq. \ref{relation:6} to obtain
\begin{align}\label{relation:8}
g( \eavec ,\mathcal L_{\partial_t} \ebvec ) 
\, -& \,
g( \ebvec ,\mathcal L_{\partial_t} \eavec ) \nonumber \\
\, \equiv&\, 2\,\gamma_{ij}\,e_{[\hat a}{}^{i}\partial_t e_{\hat b]}{}^{j}
= 0
\; .
\end{align}
Application of this result to Eq. \ref{spin:ab0:6} results in a now simpler representation of $\spinfour_{\hat a\hat b\hat0}$:
\begin{equation}\label{spin:ab0:8}
\spinfour_{\hat a\hat b\hat0}
= 
 \frac{1}{2\alpha} \bigl( 
 g( \ebvec, \Liebeta \eavec 
-
 g( \eavec, \Liebeta \ebvec ) 
\bigr) 
\; .
\end{equation}

Express the Generic Tetrad in terms of the metric basis, $\eavec = e_{\hat a}{}^i\partial_i$. The inner products of Eq. \ref{spin:ab0:8} can then be evaluated using a bilinear scalar function defined as
\begin{equation}\label{S:defn}
  S( {\bf a}, {\bf b} ) := g( {\bf a}, \Liebeta {\bf b}) - g( {\bf b}, \Liebeta {\bf a})
  \; .
\end{equation}
This reduces Eq. \ref{spin:ab0:8} to a sum of metric inner products:
\begin{equation}\label{spin:ab0:9}
\spinfour_{\hat a\hat b\hat0} 
\,= \, -\frac{1}{2\alpha} \ea^i \eb^j \, S\big( \partial_i ,  \partial_j \big) 
\; .
\end{equation}
Now use Eq. \ref{Lie} to evaluate the Lie derivative of metric basis vectors, giving
\begin{equation}\label{spin:ab0:10}
\Liebeta \partial_j = [\betavec,\partial_j] = -\partial_j \beta^i\,\partial_i \; .
\end{equation}
Project Eq. \ref{spin:ab0:10} onto the the spatial metric basis vector, $\partial_i$:
\begin{equation}\label{spin:ab0:12}
\gamma\left( \partial_i \, , \, \Liebeta \partial_j \right) 
= -\gamma_{iq} \, \partial_j \beta^q\
\; .
\end{equation}
Apply this to bilinear operator, $S$, of Eq. \ref{S:defn}:
\begin{equation}\label{spin:ab0:14}
S(\partial_i,\partial_j) =  \gamma_{j q} \partial_i \beta^q - \gamma_{i q} \partial_j \beta^q 
\; .
\end{equation}
In turn, this can be utilized in Eq. \ref{spin:ab0:9} to construct an equivalent representation of $\spinfour_{\hat a\hat b\hat0}$:
\begin{equation}\label{spin:ab0:16}
\spinfour_{\hat a\hat b\hat0} 
\,= \, -\frac{1}{2\alpha} \ea^i \eb^j \, \bigl( \gamma_{j q} \partial_i \beta^q - \gamma_{i q} \partial_j \beta^q   \bigr) 
\; .
\end{equation}
The rate of GFR accumulation given in Eq. \ref{rate:chi:1} can now be expressed in a more practical form:
\begin{equation}\label{rate:chi:2}
\boxed{
\frac{d\chi}{d\lambda}  
\,= \, -\frac{\wE}{2\alpha} \eone^i \etwo^j \, \bigl( \gamma_{j q} \partial_i \beta^q - \gamma_{i q} \partial_j \beta^q   \bigr) 
}
\; .
\end{equation}
Once again, this holds for any member of the Generic Tetrad family.

\section{The Shift Tetrad}\label{section:draggingtetrad}

The GFR rate of Eq. \ref{rate:chi:2} requires that the screen vectors be known explicitly. A first thought might be to adopt the FW frame, but it must be obtained by numerically solving Eq. \ref{FWE:spina33} for FWE. This motivates the search for different screen basis, ideally one that also offers geometric insight.

Since it is Lense-Thirring precession that causes any screen basis to rotate relative to a local FW frame, it makes sense to construct a new screen frame, $\{\ebetaonevec, \ebetatwovec\}$, in terms of the shift vector that embodies such frame dragging. We therefore make $\ebetatwovec$ proportional to $\khatvec\cross\betavec$ and refer to the result as the \emph{Shift Dyad}:
\begin{equation}
\ebetaonevec := \ebetatwovec \cross \khatvec  , \quad
\ebetatwovec := \frac{\khatvec \cross \betavec}{|\khatvec \cross \betavec|}
\; .
\end{equation}
Provided that the photon trajectory can be expressed using standard functions, the spatial wave vector, the shift vector, and ultimately the GFR rate of Eq. \ref{rate:chi:1}, can be evaluated in closed form as well.

\section{GFR Rate Formula}\label{section:GFR:anal}

Since the expression for $\spinfour_{\hat 1 \hat 2 \hat 0}$ given by Eq. \ref{spin:ab0:16} was derived for a Generic Tetrad, it holds for the Shift Tetrad as well:
\begin{equation}\label{claim:1}
\spinfour_{\grave 1 \grave 2 \grave 0} 
\,= \, -\frac{1}{2\alpha} \ebetaone^i \ebetatwo^j \, \bigl( \gamma_{j q} \partial_i \beta^q - \gamma_{i q} \partial_j \beta^q   \bigr) 
\; .
\end{equation}
Within the setting of Kerr spacetime\cite{Kerr2009} with Boyer-Lindquist coordinates\cite{BoyerLindquist1967}, we will now show that this frame is particularly attractive because metric stationarity and axisymmetry, combined with a purely azimuthal shift, implies that
\begin{equation}\label{claim:keyresult}
\spinfour_{\grave 1 \grave 2 \grave 0} = K_{\grave{1} \grave{2}} 
\; .
\end{equation}
Using Eq. \ref{K:Tetrads:v2}, this component of extrinsic curvature in the Shift Tetrad is given by
\begin{equation}\label{claim:2}
K_{\grave{1} \grave{2}} = \ebetaone^i \ebetatwo^j K_{ij} 
\; ,
\end{equation}
with metric components $K_{ij}$ defined in Eq. \ref{K:ADM}. Construct a tensor field by taking the difference between $\spinfour_{\grave 1 \grave 2 \grave 0}$ of Eq. \ref{claim:1} and the curvature component of Eq. \ref{claim:2}:
\begin{equation}\label{claim:4}
\Delta_{ij} := \gamma _{iq} \beta^q{}_{,j} - \gamma _{jq} \beta^q{}_{,i} + \Dthree_i \beta_j + \Dthree_j \beta_i 
\; .
\end{equation}
Eq. \ref{claim:keyresult} holds if $\Delta_{ij} \ebetatwo^j =  0$. This can be established immediately by directly substituting the Boyer-Lindquist expressions for spatial metric ${\boldsymbol\gamma}$ and shift vector $\betavec$. To explain \emph{why} the condition is satisfied, though, the structure of the terms on the right side of Eq. \ref{claim:4} are examined. Since $\betavec$ is only a function of $r$ and $\phi$, and because the spatial metric is diagonal in the Boyer-Lindquist setting, the first pair of terms at right in Eq. \ref{claim:4} comprise a skew-symmetric matrix of the form
\begin{equation}\label{claim:7}
\left(
\begin{array}{ccc}
 0 & 0 & A_{13} \\
 0 & 0 & A_{23} \\
 -A_{13} & -A_{23} & 0 \\
\end{array}
\right)
\; .
\end{equation}
Following the same reasoning, the second pair of terms at right in Eq. \ref{claim:4} is a symmetric version of this, 
\begin{equation}\label{claim:8}
\left(
\begin{array}{ccc}
 0 & 0 & A_{13} \\
 0 & 0 & A_{23} \\
 A_{13} & A_{23} & 0 \\
\end{array}
\right)
\; .
\end{equation}
Combining Eqs. \ref{claim:7} and \ref{claim:8}, Eq. \ref{claim:4} reduces to a matrix of the form
\begin{equation}\label{claim:9}
[\Delta] 
=
\left(
\begin{array}{ccc}
 0 & 0 & 2 A_{13} \\
 0 & 0 & 2 A_{23} \\
 0 & 0 & 0 \\
\end{array}
\right)
\; .
\end{equation}
Now turn to the second screen vector. It has no component in the $\phi $-direction because it is the result of taking the cross product of $\betavec$ with $\khatvec$, and $\betavec = \beta^\phi {\bf e}_\phi$. Therefore
\begin{equation}\label{claim:11}
\pmb{\Delta }\pmb{ }\pmb{e}_{\grave{2}} \Leftrightarrow  \left(
\begin{array}{ccc}
 0 & 0 & 2A_{13} \\[6pt]
 0 & 0 & 2A_{23} \\[6pt]
 0 & 0 & 0 \\
\end{array}
\right)\left(
\begin{array}{c}
 e_{\grave{2}}^1 \\[6pt]
 e_{\grave{2}}^2 \\[6pt]
 0 \\
\end{array}
\right) = \left(
\begin{array}{c}
 0 \\[6pt]
 0 \\[6pt]
 0 \\[6pt]
\end{array}
\right)
\; .
\end{equation}
This is sufficient to establish Eq. \ref{claim:keyresult}.

We are now in a position to construct a simple rate relation for the accumulation of GFR. Substitute Eq. \ref{claim:keyresult} into Eq. \ref{rate:chi:1}, and express the result in terms of the metric curvature components, Eq. \ref{K:Tetrads:v2}, to obtain
\begin{align}\label{rate:chi:primaryresult}
\boxed{
\frac{d\chi}{d\lambda} = \wE K_{\grave{1} \grave{2}} \equiv \wE \, e_{\grave i}{}^{p} \, e_{\grave j}{}^{q} K_{pq} 
}
\; .
\end{align}
This is our primary result, an exact, closed-form expression for the rate of GFR relative to FW frames aligned with the spatial wave vector along the photon trajectory.

\section{Shift Dyad Rotation}\label{section:elucidation}

The prescence of $K_{\grave{1} \grave{2}}$ in our rate formula can be understood by adapting an earlier definition of curvature, Eq. \ref{Kop2}:
\begin{equation}\label{Kop2again}
K_{\grave{1}\grave{2}} = -g\bigl(\ebetaonevec \, , \, (\ebetatwovec , \gradfour) \Uvec \bigr) 
\; .
\end{equation}
This makes it clear that $K_{\grave{1}\grave{2}}$ is the differential rotation of screen vector $\ebetatwovec$ towards screen vector $\ebetaonevec$ as measured along the path of the Eulerian observer. It is not the local vorticity of the hypersurface congruence (which is zero); instead it accounts for the extrinsic tilt of the Shift Triad \emph{within the slice} as the foliation deforms.  This  geometric imprint of the Lense-Thirring effect is perceived as GFR.  Although there are other screen bases that can be applied to the rate relation of Eq. \ref{rate:chi:1}, the Shift Tetrad is the simplest because it embodies a precise, one-to-one relation between dyad rotation and frame dragging. This curvature-based formulation offers an Eulerian perspective on GFR that is complementary to the more standard gravitomagnetic perception of a Lagrangian observer within the threading (1+3) split of spacetime.

\section{Implementation}\label{section:Verification}

Our primary result is the rate formula for GFR, Eq. \ref{rate:chi:primaryresult}, which can be evaluated analytically provided that the same is true for the path of photons.  We now consider two types of photon trajectories for which this is so: a closed path that remains outside the ergosphere; and a scattering trajectory that passes in and out of the ergosphere. The GFR predictions are shown to be consistent with results obtained by numerically solving both the parallel transport and Fermi-Walker equations. 

\subsection{Photon Trajectories}\label{subsection:PhotonTrajs}
 
Within the Eikonal ansatz, a large class of photon trajectories can be described with Jacobi elliptic functions by analytically solving the parallel transport dynamics of the wave vector, expressed as the derivative of position via Eq. \ref{affinedefn}\cite{Gralla_2020, Wang_2022, Lusk_PRD_2024, Lusk_Parvin_2025}. Besides the initial position and momentum, there are only two independent scalars in the associated coordinate equations: impact parameter $\psi := \ell / \varepsilon$ and Carter ratio $\eta:= Q / \varepsilon^2$. Here $\ell$ and $\varepsilon$ are the conserved scalars associated with metric isometries in the direction of its basis vectors ${\bf e}_3$ and ${\bf e}_0$, while $Q$ is the conserved (Carter) scalar that derives from the application of the Walker-Penrose Theorem\cite{Walker_Penrose_1970, Penrose1973} to a Killing-Yano 2-form\cite{Carter1968, Connors_1977, Chandrasekhar1998, Lusk_PRD_2024}. 

\begin{figure}
    \centering
        \includegraphics[width=0.35\textwidth]{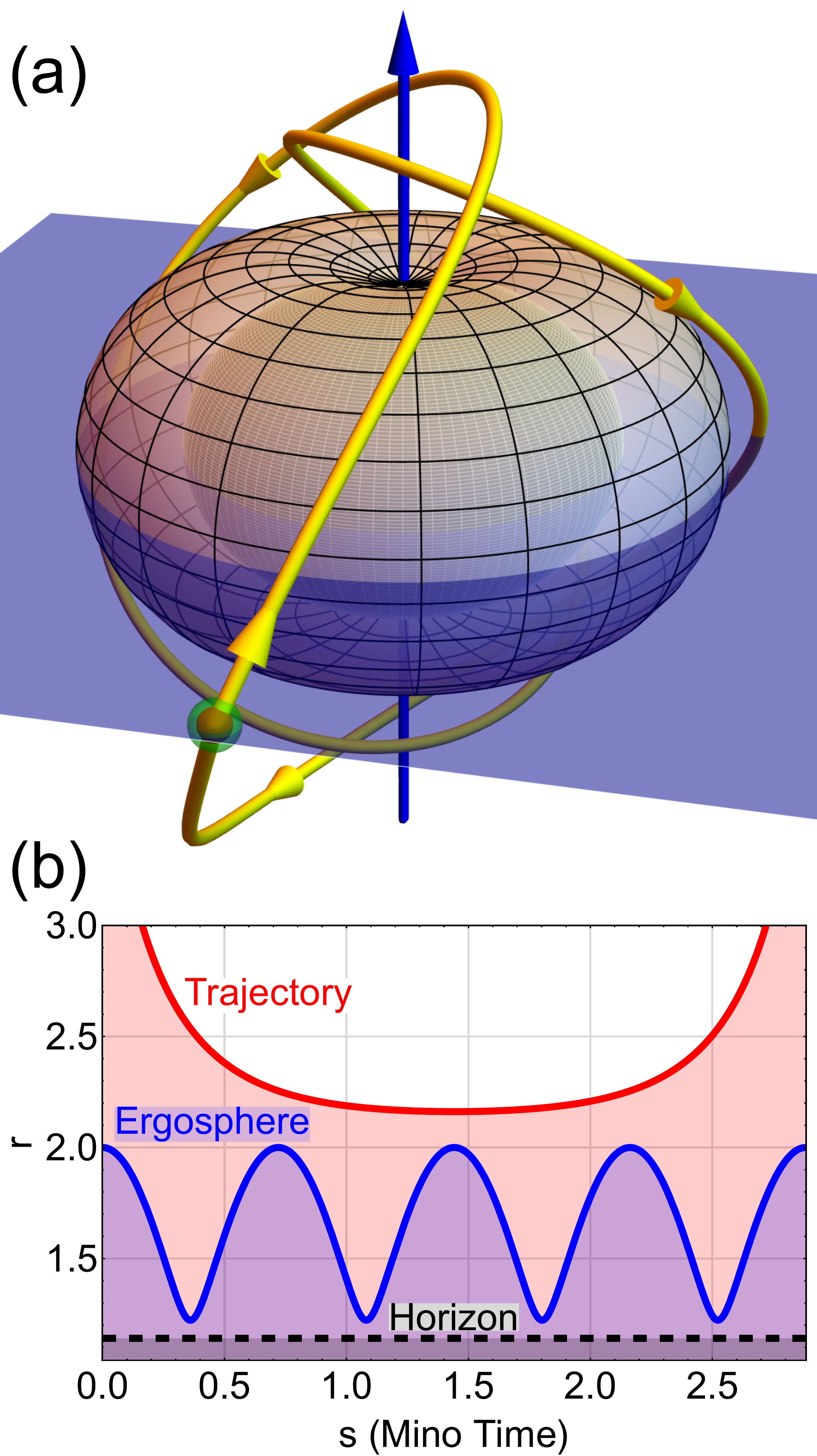}
        \caption{ \emph{Closed Trajectory Transit Outside the Ergosphere.} (a) A closed photon trajectory is shown in yellow along with the common starting (green) and stopping  (red) points. The ergosphere (white with black mesh) and outer event horizon (black with white mesh) make it clear that this trajectory stays outside the ergosphere while carrying out a non-symmetric, three-dimensional clover leaf path. (b) The radial component of the trajectory is plotted along with the associated ergosphere and event horizon radii at the same polar angle. This shows that the trajectory remains external to the ergosphere. $\psi = 0.766295, \eta = 17.9662, a = 0.99, r_i = r_f = 4$.} 
       \label{Visualize_Closed_Traj}
\end{figure}

\begin{figure}
    \centering
        \includegraphics[width=0.35\textwidth]{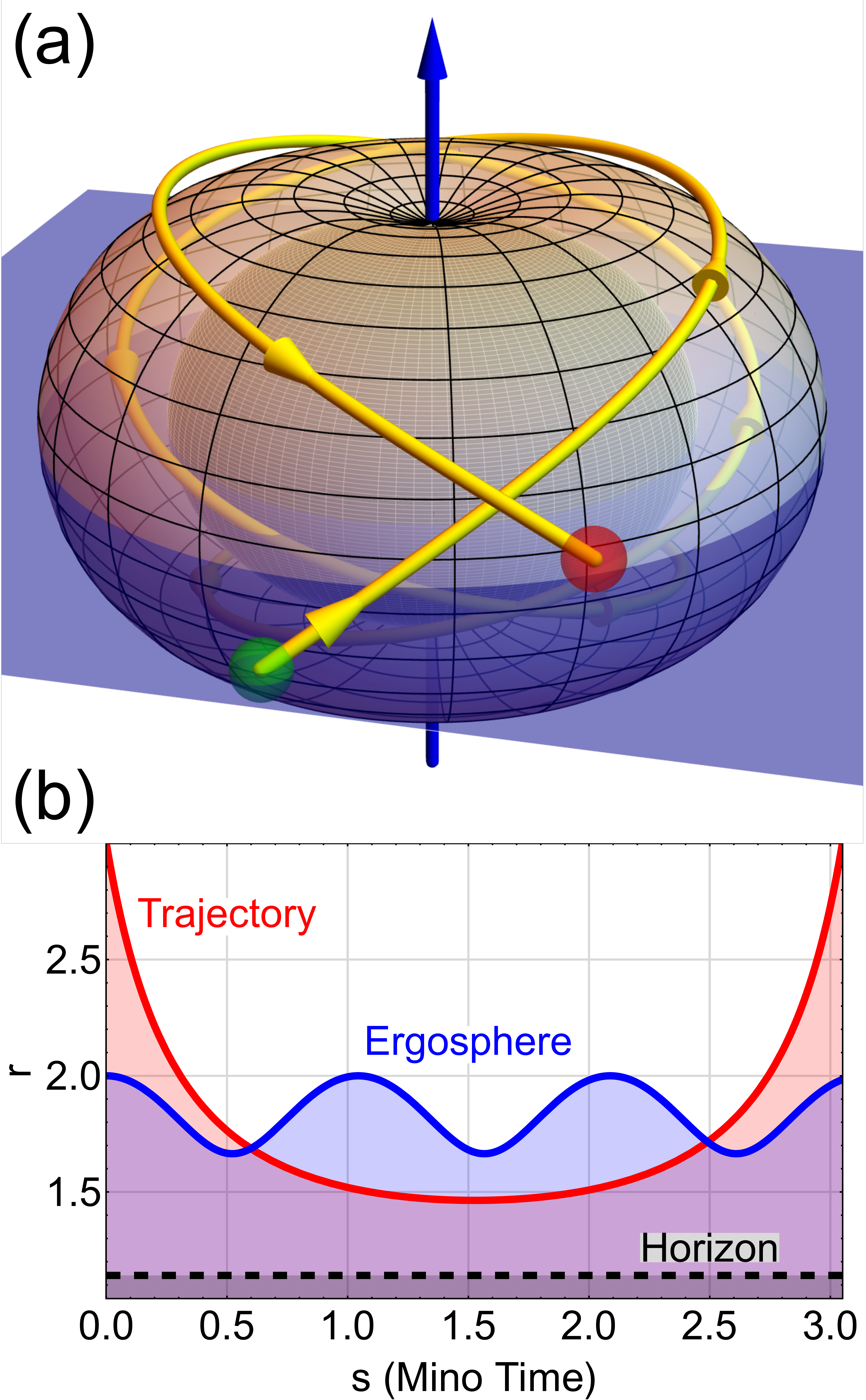}
        \caption{ \emph{Open Trajectory Transit Through the Ergosphere.} (a) An open photon trajectory is shown in yellow along with distinct starting (green) and stopping  (red) points. The ergosphere (white with black mesh) and outer event horizon (black with white mesh) make it clear that this trajectory starts and ends outside the ergosphere but transits through the ergosphere for much of the path shown. (b) The radial component of the trajectory is plotted along with the associated ergosphere and event horizon radii at the same polar angle. This shows that the trajectory pierces the ergosphere twice. $\psi = 2.05, \eta = 5, a = 0.99, r_i = r_f = 3$.} 
        \label{Visualize_Open_Traj}
\end{figure}

Analytical solutions are made possible by replacing rates measured using affine parameter, $\lambda$, with those expressed in terms Mino time, $s$\cite{Mino_2003}.  The two parameterizations are related by
\begin{equation}\label{Mino}
\frac{\dd \lambda}{\dd s} \equiv \Sigma :=  r^2 + a^2 \cos^2\theta
\; ,
\end{equation}
which decouples the coordinate equations of motion.
This formalism has been used previously to quantify polarization holonomy within the threading split of spacetime, where GFR is measured by a Lagrangian ghost\cite{Lusk_PRD_2024, Lusk_Parvin_2025}. Two such trajectories shown in Figs. \ref{Visualize_Closed_Traj} and \ref{Visualize_Open_Traj}.

%
%

%
%

\subsection{GFR Predictions}\label{subsection:chi:analytical}

\begin{figure}[t]
	\begin{center}
		\includegraphics[width=0.35\textwidth]{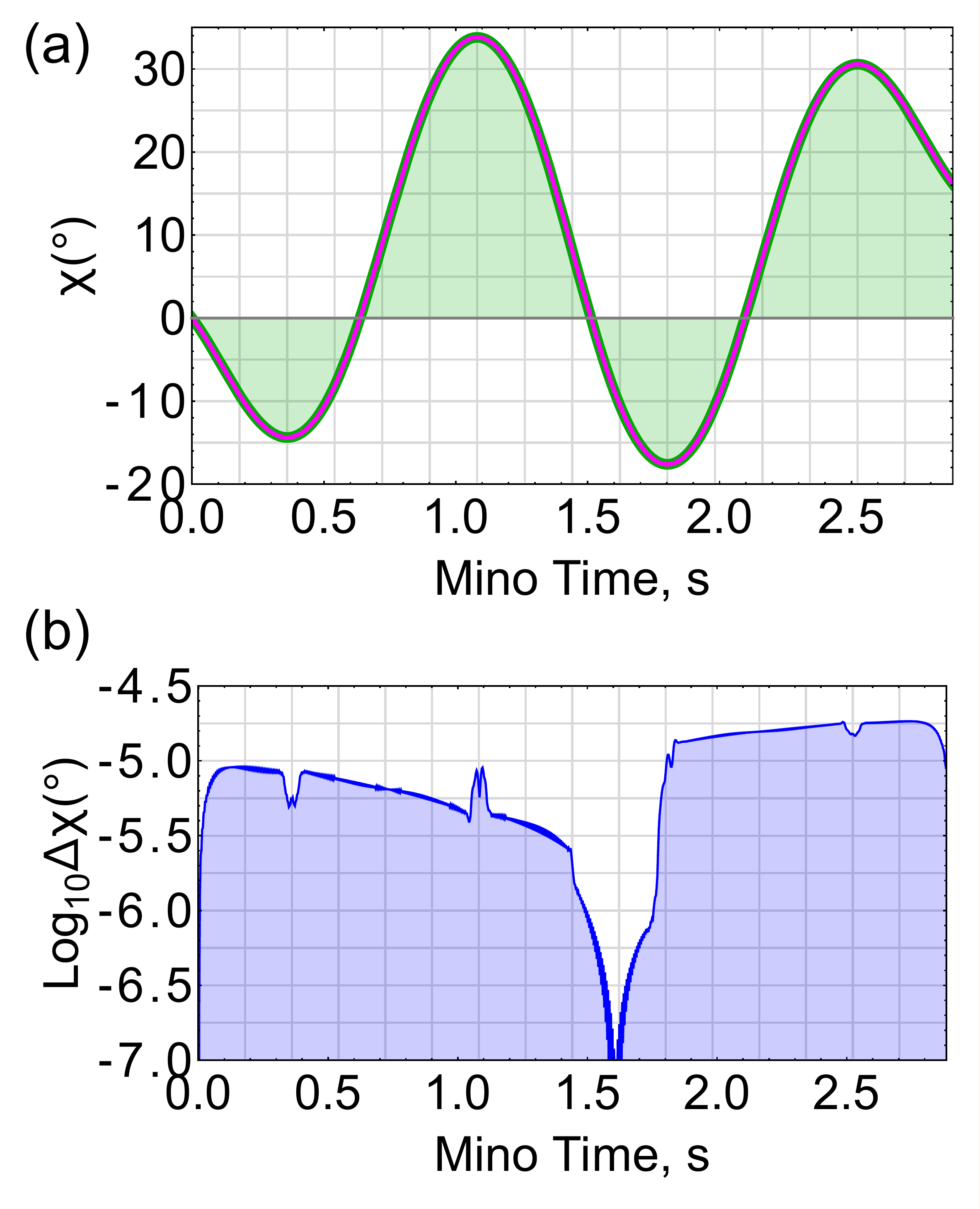}
	\end{center}
	\caption{ \emph{GFR for Closed-Loop Trajectory of Fig. \ref{Visualize_Closed_Traj}.} (a) Analytical prediction of GFR (green) along with numerical result (magenta), (b) Discrepancy $\Delta\chi(s)$, Eq. \ref{Deltachi}, is plotted on a log scale and is on the order of $10^{-5}$ \degree.} 
	\label{Verification_Closed_Traj}
\end{figure}

\begin{figure}[t]
	\begin{center}
		\includegraphics[width=0.35\textwidth]{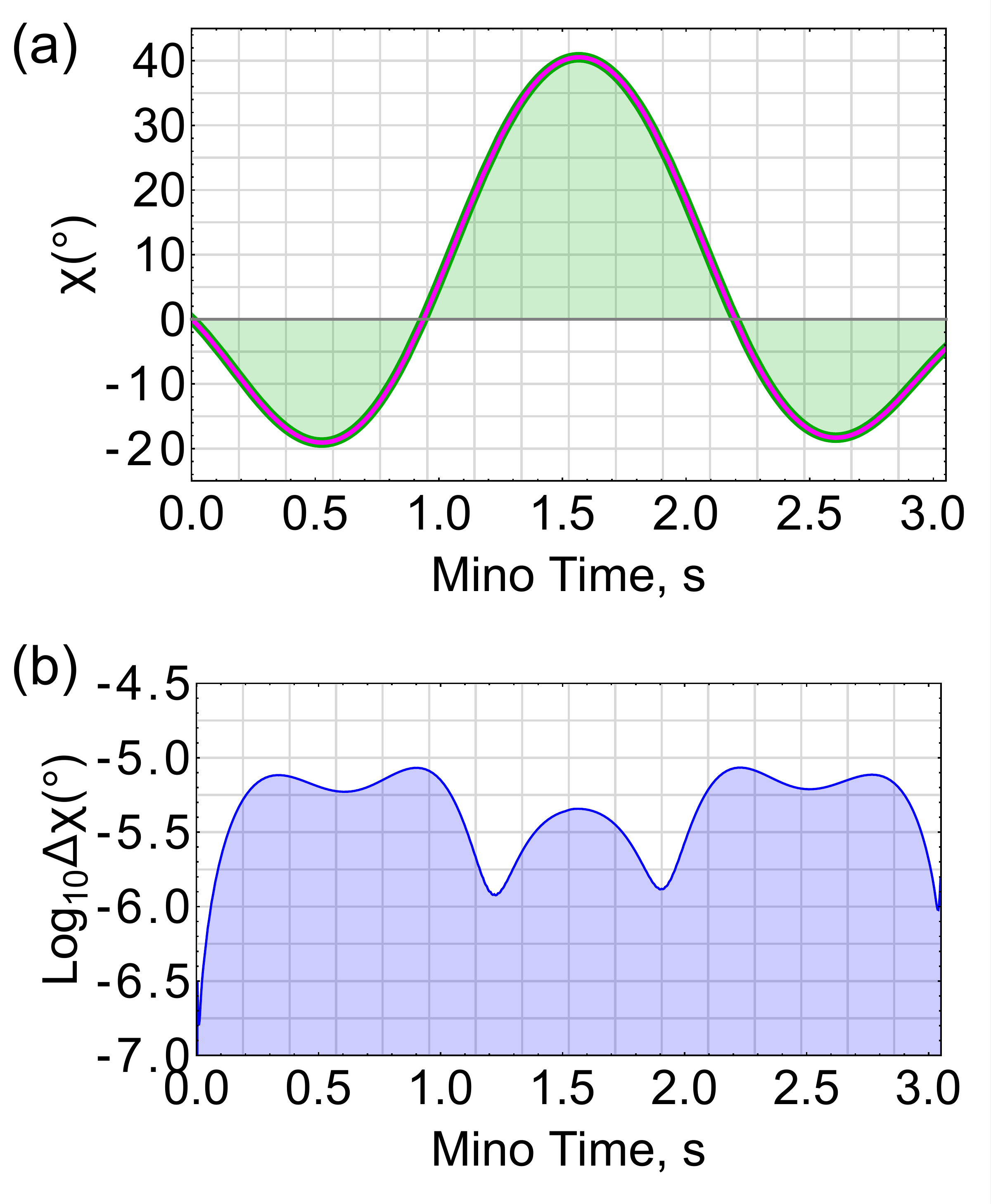}
	\end{center}
	\caption{ \emph{GFR for Open Trajectory of Fig. \ref{Visualize_Open_Traj}.} (a) Analytical prediction of GFR (green) along with numerical result (magenta), (b) Discrepancy $\Delta\chi(s)$, Eq. \ref{Deltachi}, is plotted on a log scale and is on the order of $10^{-5}$ \degree.} 
	\label{Verification_Open_Traj}
\end{figure}

The rate of accumulation of GFR, Eq. \ref{rate:chi:primaryresult}, is implemented for the photon trajectories of Figs. \ref{Visualize_Closed_Traj} and \ref{Visualize_Open_Traj}. Its integrated accumulation is plotted in panel (a) of Figs. \ref{Verification_Closed_Traj} and \ref{Verification_Open_Traj}. The ability to produce such predictions is our principal product.
%
%
%
%
%

%
%
%

A numerical procedure can be used as a consistency check on these predictions. Prior to any gauge fixing, the free polarization vector, $\polfourfreevec$, is parallel transported in spacetime, and this can be numerically evaluated by solving Eq. \ref{pol:evol}. The gauge is then fixed using Eq. \ref{gauge:function}.  An orthonormal pair of linear polarizations, $\polAthreefixed$ and $\polBthreefixed$, are constructed using Eq. \ref{gauge:polfixed}.  These, in turn, are used to generate a circular polarization,
\begin{equation}\label{circPX:defn}
\polthreecircvec := \frac{e^{i \chi}}{\sqrt{2}} ( \polAthreefixed - i  \, \polBthreefixed )
\; .
\end{equation}
The initial circular polarization can also be Fermi-Walker transported by numerically solving Eq. \ref{FWE:spina33} to obtain $\mFWE$. 

GFR is then defined as the difference in the phase between $\polthreecircvec$ and $\mFWE$:
\begin{equation}\label{chi:numerical}
\chinumerical\!\!\!(s) := \arg \gamma\bigl( \polthreecircvec\!\!(s)  \, , \, \mFWE\!\!(s)  \bigr)
\; .
\end{equation}

The equivalence of analytical predictions and numerical results is established by generating a discrepancy metric of 
\begin{equation}\label{Deltachi}
\Delta\chi(s) := \left| \chinumerical\!\!\!(s) \; - \!\chianal\!\!\!(s) \right| 
\; .
\end{equation}
This is plotted in panel (b) of Figs. \ref{Verification_Closed_Traj} and \ref{Verification_Open_Traj}. The discrepancy is on the order of $10^{-5 \,\circ}$, a level attributable to the numerical methods used, so the numerical method is deemed consistent with our analytical formula.

\subsection{Single-Observer Measurement of GFR}

The measurement of GFR within the slicing split of spacetime relies on multiple Eulerian observers with positions that coincide with the photon trajectory. It may well be possible, though, for a single Eulerian observer to compare a measured polarization holonomy with theoretical prediction. The Eulerian (hypersurface) angular velocity,  
\begin{equation}
\Omega_{\rm E} =\frac{d\phi}{dt} \equiv \frac{-g_{t\phi}}{g_{\phi\phi}}  \equiv -\,\beta^\phi 
\, ,
\end{equation}
can be converted to a rate of change with respect to Mino time,
\begin{equation}
\Omega_{\rm E, Mino} =\Omega_{\rm E} \frac{dt}{ds} \equiv -\,\beta^\phi \frac{dt}{ds}
\, ,
\end{equation}
where the temporal rate is known analytically\cite{Gralla_2020, Lusk_PRD_2024, Lusk_Parvin_2025}. Since this is not a function of azimuthal angle, the Mino time required for an Eulerian observer to return to their initial spatial coordinates is
\begin{equation}
s_{\rm E} = \frac{2\pi}{\Omega_{\rm E, Mino}}
\, .
\end{equation}
In the closed trajectory of Fig. \ref{Visualize_Closed_Traj}, $s_{\rm E} = 7.28231$ while the photon trajectory plays out over $s_\gamma = 2.88332$, during which the photon makes three transits about the singularity. The ratio of these is not an integer, but a plausible conjecture is that the setting could be modified so that it is. In that case, a single Eulerian observer would be present for both the emission and capture of a given photon. The idea can be generalized, as well, so that the observer and photon meet at a time-synchronized location that is distinct from the emission point. 

As an aside, it is also true that a \emph{stationary} observer, located at a common location of source/receiver could produce a weaker, model-based verification of the slicing GFR prediction if equipped with the rate of rotation of the singularity.

\section{Conclusions}\label{section:conclusions}

A continuum of Eulerian observers along the trajectory of a photon can individually measure an instantaneous rate of polarization rotation with respect to a local frame of reference. The reference screen dyad is governed by Fermi-Walker rotation about the local spatial wave vector instead of the photon trajectory. Path integration then quantifies the accumulation of GFR. Within the ADM split of Kerr spacetime, these Eulerian spotters have zero angular momentum and, from their perspective, the rotation of the polarization of light relative to FW frames is due to the frame dragging deformation of the hypersurface, completely characterized by its extrinsic curvature. When applied to a specially engineered, closed trajectory, it should be possible for a single Eulerian observer to compare a measured polarization holonomy with theoretical prediction. 

The rate of GFR is formulated using a Shift Dyad, $\{\ebetaonevec, \ebetatwovec\}$, for which $\ebetatwovec \,\,{\scriptstyle \propto}\, \,\khatvec\cross\betavec$.  The spatial gradient of the shift vector then has a component which produces a differential rotation of the screen basis about the foliation normal.  This distills into an exact, closed-form rate of GFR proportional to the $K_{\grave 1 \grave 2}$--component of the extrinsic curvature in our bespoke frame. Such a curvature-based formulation offers an Eulerian perspective on GFR that is complementary to the more standard gravitomagnetic perception of a Lagrangian ghost within the threading (1+3) split of spacetime. 

The new rate of GFR accumulation has been formulated using a specific polarization gauge wherein spatial polarization is orthogonal to the spatial wave vector.  The accumulated GFR is invariant with respect to gauge, though. This has been numerically checked by adopting a gauge for which the spatial polarization is orthogonal to the photon trajectory instead of the spatial wave vector.  It is worth emphasizing that the GFR perceived by an Eulerian observer is not the same as that measured by a Lagrangian observer. This is because the reference triads are aligned with the spatial wave vector and spatial photon trajectory, respectively.

The Eulerian formulation of GFR does not suffer from the mathematical singularity at the ergosphere present for Boyer-Lindquist coordinates in the threading split. These physically intuitive coordinates can therefore be used to analytically produce and scrutinize GFR predictions for transits of light that pierce the ergosphere.

\bibliographystyle{prsty} 	
\bibliography{GFR_Slicing}

\end{document}